\documentclass[twocolumn,epjc3]{svjour3}  
\RequirePackage[T1]{fontenc}
\smartqed  % flush right qed marks, e.g. at end of proof
\RequirePackage{graphicx}
\RequirePackage{mathptmx}      % use Times fonts if available on your TeX system
\RequirePackage{flushend}
\RequirePackage{latexsym}
\RequirePackage[colorlinks,citecolor=blue,urlcolor=blue,linkcolor=blue]{hyperref}
\RequirePackage{amsmath,amssymb}
\newcommand{\e}{{\sf e}}
\journalname{Eur. Phys. J. C}
\begin{document}

\title{Dispersive analysis for $\eta\to \gamma\gamma^\star$%\thanksref{t1}
}
%\subtitle{Do you have a subtitle?\\ If so, write it here}

%\titlerunning{Short form of title}        % if too long for running head

\author{C.~Hanhart\thanksref{e1,addrJ1,addrJ2,addrJ3}
        \and
        A.~Kup{\'s}{\'c}\thanksref{e2,addrK1,addrK2} 
        \and
        U.-G.~Mei{\ss}ner\thanksref{e3,addrJ1,addrJ2,addrJ3,addrBN1,addrBN2}
        \and
        F.~Stollenwerk\thanksref{e4,addrJ1,addrH}
        \and
        A.~Wirzba\thanksref{e5,addrJ1,addrJ2,addrJ3}
}

%\thankstext{t1}{Grants or other notes
%about the article that should go on the front page should be
%placed here. General acknowledgments should be placed at the end of the article.
\thankstext{e1}{e-mail: c.hanhart@fz-juelich.de}
\thankstext{e2}{e-mail: Andrzej.Kupsc@physics.uu.se}
\thankstext{e3}{e-mail: meissner@hiskp.uni-bonn.de}
\thankstext{e4}{e-mail: felix.stollenwerk@physik.hu-berlin.de}
\thankstext{e5}{e-mail: a.wirzba@fz-juelich.de}

%\authorrunning{Short form of author list} % if too long for running head

\institute{Institut  f\"{u}r Kernphysik (Theorie), Forschungszentrum J\"ulich,  
             D-52425 J\"{u}lich, Germany  \label{addrJ1}
             \and
               Institute for Advanced Simulation, Forschungszentrum J\"ulich,  
               D-52425 J\"{u}lich, Germany  \label{addrJ2}
             \and
            J\"ulich Center for Hadron Physics, Forschungszentrum J\"ulich,  
            D-52425 J\"{u}lich, Germany  \label{addrJ3}
             \and
           Division of Nuclear Physics, Department of Physics and Astronomy
           Uppsala University,  Box 516, 75120 Uppsala, Sweden  \label{addrK1}
           \and
           High Energy Physics Department,
           National Centre for Nuclear Research, 
           ul.\ Hoza 69, 00-681 Warsaw, Poland  \label{addrK2}
           \and
            Helmholtz-Institut f\"ur Strahlen- und Kernphysik, 
            Universit\"at Bonn, D-53115  Bonn, Germany \label{addrBN1}
           \and
           Bethe Center for Theoretical Physics,
           Universit\"at Bonn, D-53115  Bonn, Germany \label{addrBN2}
           \and
           \emph{Present Address:} 
           Institut f\"ur Physik, Humboldt-Universit\"at zu Berlin, 
            Newtonstr.~15, D-12489 Berlin, Germany \label{addrH}
}

\date{Received: date / Accepted: date}
% The correct dates will be entered by the editor

\maketitle

%%%%%%%%%%%%%%%%%%%%%%%%%%%%%%%%%%%%%%%%%%%%%%%%%%
%%%%%%%%%%%%%%%%%%%%%%%%%%%%%%%%%%%%%%%%%%%%%%%%%%

\begin{abstract}
A dispersion integral is derived that connects data on $\eta\to \pi^+\pi^-\gamma$
to the $\eta\to \gamma\gamma^*$ transition form factor. A detailed
analysis of the uncertainties is provided. 
We find for the slope of the  $\eta$ transition form factor at the origin  
 $b_\eta = \left(2.05 \ \mbox{}^{+0.22}_{-0.10} \right)\text{GeV}^{-2}$.    
Using an additional, plausible assumption, one finds for the corresponding slope of the
$\eta'$ transition form factor, $b_{\eta'} = \left(1.53\   \mbox{}^{+0.15}_{-0.08} \right)\text{GeV}^{-2}$.
Both values are
consistent with all recent
data, but differ from some previous theoretical analyses.
\keywords{Form factors \and Dispersion integral \and Vector meson dominance}
% \PACS{PACS code1 \and PACS code2 \and more}
% \subclass{MSC code1 \and MSC code2 \and more}
\end{abstract}

%%%%%%%%%%%%%%%%%%%%%%%%%%%%%%%%%%%%%%%%%%%%%%%%%%
%%%%%%%%%%%%%%%%%%%%%%%%%%%%%%%%%%%%%%%%%%%%%%%%%%

\section{Introduction}
\label{intro}

Transition form factors contain important information about the properties of 
the decaying particles. Additional interest into meson decays with one or two virtual
photons in the final state comes from the fact that the theoretical uncertainty
for the Standard Model calculations for $(g-2)$ of the muon will soon be completely
dominated by the hadronic light-by-light amplitudes, where they appear as
sub-amplitudes---for a recent discussion of this issue see Refs.~\cite{krakau,trento}.

In this work, using dispersion theory,
the connection between the radiative decays
$\eta\to \pi^+\pi^-\gamma$  and  $\eta'\to \pi^+\pi^-\gamma$ and the isovector contributions
of the 
form factors $\eta\to\gamma\gamma^*$ and  $\eta'\to\gamma\gamma^*$
is exploited in a model-independent way. This is possible, because the amplitude of the former 
decays can be parametrized in terms of the pion vector form factor, $F_V(Q^2)$, and a 
low-order polynomial~\cite{Stollenwerk}, 
since $F_V(Q^2)$ as well as  the radiative decay amplitudes 
$\eta\to \pi\pi\gamma$  and  $\eta'\to \pi\pi\gamma$ 
share, at least in the low-energy regime,  the same right-hand cut.
Therefore the vector form factor and the decay amplitudes
must agree up to a function that is free
of a right-hand cut and therefore varies only smoothly with $Q^2$---the
invariant mass squared of the pion pair. It was therefore proposed
to parametrize the differential decay widths for $\eta\to \pi\pi\gamma$ (and analogously for  $\eta' \to \pi\pi\gamma$) as
\begin{equation}
\label{eq:main}
\frac{d\Gamma^{\eta}_{\pi\pi\gamma} }{dQ^2} = \left | A_{\pi\pi\gamma}^{\eta} \,
P(Q^2)F_V(Q^2)\right|^2 \Gamma_0(Q^2)\ ,
\end{equation}
where the normalization parameter $A_{\pi\pi\gamma}^{\eta}$, which is determined by the empirical value
of the partial
decay width \cite{pdg:2012}, has the dimension of ${\rm mass}^{-3}$. The function
\[
  \Gamma_0(Q^2) = \frac{1}{3 \cdot 2^{11} \cdot \pi^3 \,m_P^3} 
  \left(m_P^2-Q^2 \right)^3  Q^2 \,\sigma_\pi(Q^2)^3
\]
collects phase-space terms and the kinematics of the absolute square of the
simplest gauge invariant matrix element (for point-particles).  The $\pi\pi$-two-body phase space reads
$\sigma_\pi(Q^2)=\sqrt{1 - 4m_\pi^2/Q^2}$, where  $m_P$ ($m_\pi$) denotes the mass of
the decaying particle (charged pion).

In order to fit  the spectral shape
of the radiative  $\eta$ \cite{WASAeta2pipiga}
and $\eta'$ decays \cite{Crystal_Barrel},  a linear polynomial was sufficient for specifying the function 
$P(Q^2)$~\cite{Stollenwerk}.
In addition,
the slope extracted from the two fits were consistent within uncertainties---a finding that can be understood
using arguments from large $N_c$ chiral perturbation theory. 
We may therefore write
\begin{equation}
P(Q^2) = 1 + \alpha Q^2 \ ,
\label{alphadef}
\end{equation}
identifying $\alpha$ as a fundamental parameter to characterize the decays
$\eta\to \pi\pi\gamma$  and  $\eta'\to \pi\pi\gamma$.

In this paper we will use the findings of Ref.\,\cite{Stollenwerk} to 
predict the $\eta/\eta'\to \gamma\gamma^\ast$
transition form factor and its slope at the origin with the help of dispersion integral techniques.
In the rest frame of the $\eta$ meson,  say, 
the transition amplitude for $\eta\to \gamma \gamma^*$ may be decomposed as
\begin{eqnarray}
 \mathcal A^{rm} (Q^2) &=&  
  \mathcal A_1^{rm} (Q^2) + \mathcal A_0^{rm} (Q^2) \nonumber \\
 &=&\mathcal A^{rm} (0) + \Delta\mathcal A_1^{rm} (Q^2) + \Delta\mathcal A_0^{rm} (Q^2) \ ,
\label{Adecomp}
\end{eqnarray}
where $r$ and $m$ are the spatial indices of the polarization vectors of the two outgoing photons
and $\mathcal A_1^{rm} (Q^2)$ and  $\mathcal A_0^{rm} (Q^2)$
label
the isovector and isoscalar contributions to the transition amplitude, respectively.
The  $Q^2$ dependence  of the latter  are isolated in
$\Delta\mathcal A_1^{rm} (Q^2)$ and  $\Delta\mathcal A_0^{rm} (Q^2)$, which both
are  normalized to zero at $Q^2=0$.
Furthermore there is the double-on-shell amplitude
\begin{equation}
  \mathcal A^{rm} (0)\equiv \mathcal A^{rm} (\eta \to \gamma \gamma)
 = {A_{\gamma\gamma}^\eta} m_\eta \epsilon^{mrb} p_\gamma^b 
 \label{Armzero}
\end{equation}
in terms of the three-momentum of the on-shell photon, $\vec{p}_\gamma$, defined in the $\eta$ rest frame,
and of the mass of the decaying pseudo-scalar, $m_\eta$. The quantity 
\begin{equation}
A_{\gamma\gamma}^\eta \equiv \sqrt{\Gamma_{\gamma\gamma}^\eta \,  64\pi/m_\eta^3}  
\label{agamgamdef}
\end{equation}
is specified by
the $\eta\to\gamma\gamma$  partial decay width $\Gamma^\eta_{\gamma\gamma}$~\cite{pdg:2012}.

In the following, we will make model-independent predictions 
for  $\Delta\mathcal A_1^{rm} (Q^2)$ based on a dispersion integral that only needs
$P(Q^2)$ as well as $F_V(Q^2)$ as input. This analysis in principle requires knowledge about 
these quantities up to infinite values of $Q^2$; however, as we will demonstrate in the next sections,
the relevant dispersion integral is largely saturated in a regime where we do control the
input. In addition, the uncertainties from the kinematic regions where, {\it e.g.\/}, the function $P(Q^2)$ 
is not well known, can be reliably estimated.  
However, we still need model assumptions,
in particular vector-meson dominance (VMD), in order to constrain 
$\Delta\mathcal A_0^{rm} (Q^2)$. Nevertheless,  we will show that
we can even deduce the isoscalar contribution 
$\Delta\mathcal A_0^{rm} (Q^2)$ to the amplitude directly from data by only assuming  that it is the dominated by narrow 
$\omega$ and $\phi$ meson resonances.
In this way
a nearly model-independent evaluation of  the {\em complete} 
transition amplitude is provided,
valid for small values of~$Q^2$.

The paper is structured as follows: in the next section we will update the analysis of Ref.~\cite{Stollenwerk}
and also discuss the behavior of $P(Q^2)$  in the complete region $4m_\pi^2\leq Q^2\leq 1\,\text{GeV}^2$. In the subsequent section the dispersion
integral for the iso\-vector part of the $\eta/\eta'\to \gamma\gamma^*$ transition form factor and
its slope is derived,  followed
by a discussion of a model for the isoscalar counter part. We close with a presentation of the results 
and a summary. A comparison with the vector-meson dominance approximation is relegated to
the appendix.

%%%%%%%%%%%%%%%%%%%%%%%%%%%%%%%%%%%%%%%%%%%%%%%%%%
%%%%%%%%%%%%%%%%%%%%%%%%%%%%%%%%%%%%%%%%%%%%%%%%%%

\section{Remarks on the radiative decays of $\eta$ and $\eta'$}
\label{sec:eta2pipigam}

In this section we update the results of Ref.~\cite {Stollenwerk} since new data were published
in the meantime~\cite{KLOE_2013}. In addition, we provide arguments why $P(Q^2)$ can be assumed linear in
the whole range of $4m_\pi^2\leq Q^2\leq 1\,\text{GeV}^2$.

The discontinuity relation for the pion vector form factor gives
\begin{equation}
{\rm Im}(F_V(Q^2))=\sigma_\pi(Q^2) T_p^*(Q^2)F_V(Q^2) \,\Theta(Q^2-4m_\pi^2) \ ,
\label{imf}
\end{equation}
where $\Theta(\ldots)$ is the Heaviside
step function and
$T_p(Q^2)$ denotes the $\pi\pi$ elastic scattering amplitude in the $p$-wave that
may be expressed via the corresponding phase shift $\delta_p(Q^2)$ as
\begin{equation}
T_p(Q^2) = \frac1{\sigma_\pi(Q^2)}\sin (\delta_p(Q^2))\exp(i\delta_p(Q^2)) \ .
\end{equation}
Below we use the phase shifts from the  analysis of Ref.~\cite{madrid}.

If one assumes that the two-pion interactions are elastic up to infinite energies, the 
dispersion integral that emerges from Eq.~(\ref{imf}) can be solved analytically yielding
the celebrated Omn\`es function
\begin{equation}
 \Omega(Q^2) 
    = \exp\left(\frac{Q^2}{\pi}\int_{4 m_\pi^2}^\infty \frac{ds}{s}\,\frac{\delta_p(s)}{s-Q^2-i\epsilon}\right)  \ .
\end{equation}
Since any function that is multiplied to  $F_V(Q^2)$ and that is real on the right-hand cut
does not spoil Eq.~(\ref{imf}),
one may write in general
\begin{equation}
F_V(Q^2)=R(Q^2)\Omega(Q^2) \ .
\end{equation}
An identical derivation leads us to the analogous expression for the amplitudes for
the radiative decays of $\eta$ and $\eta'$, {\it e.g.\/},
\begin{equation}
\mathcal{A}_{\pi \pi\gamma}^\eta(Q^2)=A_{\pi \pi\gamma}^\eta P_\Omega(Q^2)\Omega(Q^2) \ ,
\end{equation}
where, using $P_\Omega(0)=1$ and $\Omega(0)=1$, $\mathcal{A}_{\pi \pi\gamma}^\eta(0)
={A}_{\pi \pi\gamma}^\eta$.

%%%%%%%%%%%%%%%%%%%%%%%%%%%%%%%%%%%%%%%%%%%%%%%%%%
\begin{figure}[t]
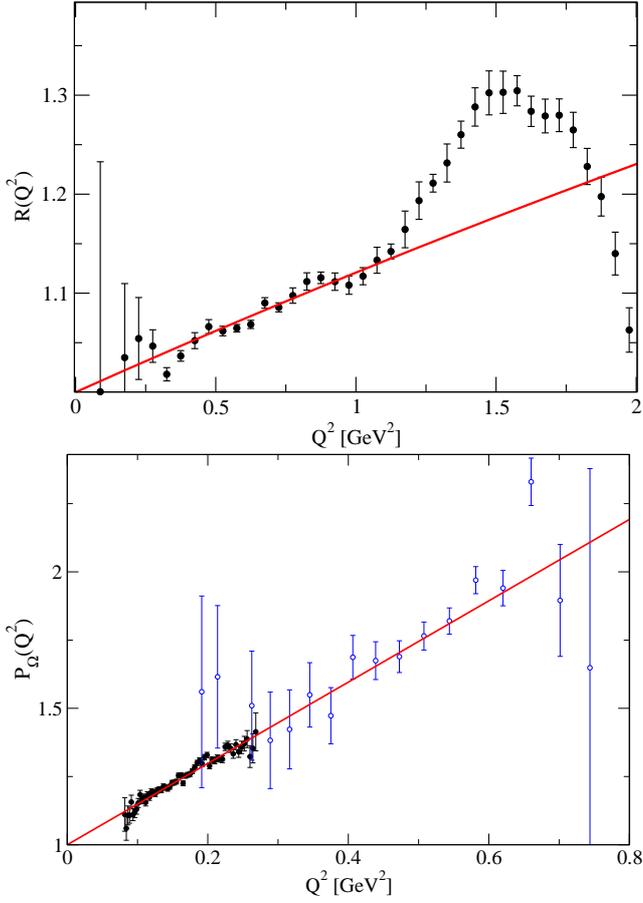
 % Figure 1
 \begin{minipage}{\columnwidth}
 \centering
   \includegraphics[width=\textwidth]{tauFFoveromnes_zoom.eps}
   \includegraphics[width=\textwidth]{amplitudefit_KetapoO.eps}
   \end{minipage}
   \caption{Upper panel: the function $R(Q^2)=F_V(Q^2)/\Omega(Q^2)$, where
   data from $\tau$ decays from Ref.~\cite{belle} were used for  the pion vector form factor. The (red) line
   denotes a linear fit to the data in the kinematic regime from threshold to $s= 1\,\text{GeV}^2$.
   Lower panel: the function $P_\Omega(Q^2)$ for radiative decays of the $\eta$---solid
   symbols from Ref.~\cite{KLOE_2013}---and the $\eta'$---open symbols from Ref.~\cite{Crystal_Barrel}.
   The (red) line denotes a linear fit to the $\eta$ data. \label{fig:overOmnes}}
\end{figure}
%%%%%%%%%%%%%%%%%%%%%%%%%%%%%%%%%%%%%%%%%%%%%%%%%%

In Figure~\ref{fig:overOmnes} we show the $Q^2$ dependence of $R(Q^2)$ (upper panel) and $P_\Omega(Q^2)$ (lower panel), the latter
for $\eta$ (solid symbols) as well as $\eta'$ (open symbols)  decays. As one can see, $R(Q^2)$ is perfectly linear 
for $Q^2<1\,\text{GeV}^2$. For larger values of the $\pi\pi$ invariant mass squared one finds clear deviations
from linearity---in this case caused by the $\rho'$~\cite{newff}, the first radial excitation of the $\rho$-meson.
The lower panel demonstrates that $P_\Omega(Q^2)$  is linear within
the experimental uncertainties in the
full range kinematically accessible---although the data for $\eta'$ clearly call for improvement. 
The straight line in the figure is a fit to the $\eta$ data, which demonstrates that the 
slope of the $\eta'$ spectrum is consistent with that of the $\eta$---this observation will be exploited below.

Whereas $F_V(Q^2)$ does not have a left-hand cut, the decay amplitudes for the
radiative decays of $\eta$ and $\eta'$  have one.
Since the transition $\eta^{(')}\to 3\pi$ is suppressed --- it violates the isospin symmetry ---
the leading singularity in both
cases is driven by the same $\pi\pi\eta$ intermediate state followed by $\pi\eta\to\pi\gamma$. 
However, 
it is strongly suppressed~\cite{Stollenwerk}: on the one hand for kinematical reasons, since the particle pairs in 
the $t$-channel
have to be (at least) in a relative $p$-wave to allow the transition $\pi\eta\to\pi\gamma$ to happen,  
on the other hand for dynamical reasons, since the $p$-wave $\pi\eta$ interaction  starts
only at next-to-leading order in the chiral expansion \cite{BKM_91,bastian}.
It is therefore justified to neglect it---an assumption supported by the strict linearity of $P_\Omega(Q^2)$ 
demonstrated above. Then, analogous to $F_V(Q^2)$,
 also the ratios of the $\eta$ and $\eta'$ amplitudes with respect to the Omn\`es function should be linear up 
 to  about $1\, \text{GeV}^2$.  At least up to $Q^2=m_{\eta'}^2$ with $m_{\eta'}$ the $\eta'$ mass, this can be checked experimentally
 once better data are available for the $\eta'$ radiative decays---those data should be expected 
 from BES-III~\cite{BESIII} and  CLAS~\cite{Primenet}
in the near future.
 For energies above  $1\, \text{GeV}$,
some influence from the higher $\rho$ resonances should be expected. 
  In the next section a dispersion integral is derived that allows us, using mainly the input
 described in this section, to calculate $\Delta {\cal A}_1^{rm}(Q^2)$ --- the isovector
 contribution to the slope of the $\eta\to \gamma\gamma^*$ form factor,
 defined in Eq.~(\ref{Adecomp}).

As outlined above, for $Q^2$ values up to 1 GeV$^2$ the $\eta\to\pi\pi\gamma$ transition amplitude is completely
fixed by the parameter $\alpha$ and the pion vector form factor. We here use for $\alpha$ 
the value given in Ref.~\cite{KLOE_2013},
\begin{equation}
   \alpha = (1.32\pm 0.13) \, \text{GeV}^{-2} \ .
\label{alphaval}
\end{equation}
The uncertainty contains the statistical as well as the systematic uncertainty from
the data as well as the theoretical uncertainty quoted in Ref.~\cite{Stollenwerk}.

Below we will need the transition amplitude also for larger values of $s$.
As a consistency check we confirmed that we reproduce the above value for $\alpha$ from
our own fit to the data of Ref.~\cite{KLOE_2013} using
 the full vector form factor, $F_V(Q^2)_{e^+e^-}$ of Ref.~\cite{newff} as input. 
It includes the effect
of isospin violation from $\gamma$-$\rho$ mixing ({\it cf.\/} Ref.~\cite{freds}) as well
as $\rho$-$\omega$ and  $\rho$-$\phi$ mixing and the effect of the first two excited states, $\rho'$
and $\rho''$. Clearly, the impact of the higher resonances as well as the mixing with
isoscalar vector states may depend on the reaction
channel, since there is no reason to expect their effects to be equal in $\eta$ radiative decays to
those found in the $e^+e^-$ reaction. Therefore in our analysis we also 
used  an alternative form-factor
parameterization to control the theoretical uncertainty: namely
one that is extracted from $\tau$ decays, $F_V(Q^2)_{\tau}$, and therefore does not contain any mixing with 
$\omega$, $\phi$ or $\gamma$. 
The spread in the results from using those two form factors is included in the systematic uncertainty
reported below.

%%%%%%%%%%%%%%%%%%%%%%%%%%%%%%%%%%%%%%%%%%%%%%%%%%
%%%%%%%%%%%%%%%%%%%%%%%%%%%%%%%%%%%%%%%%%%%%%%%%%%

\section{$\eta \to \gamma \gamma^\star$: dispersion relation} 
\label{sec:dispersion}

\begin{figure}   % Figure 2
\begin{minipage}{\columnwidth}
\centering
	\includegraphics[width=0.75\textwidth]{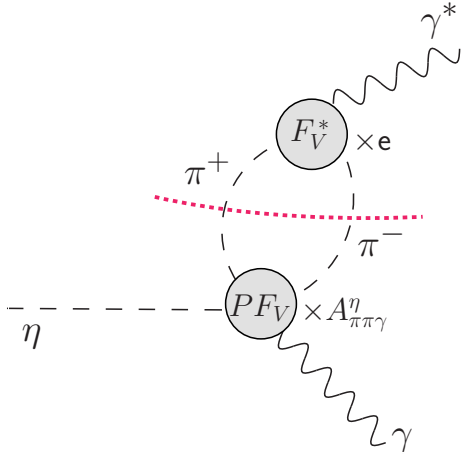}	
\end{minipage}	
	\caption{The isovector part of the $\eta\to \gamma\gamma^\ast$ decay amplitude driven by the on-shell
	two-pion intermediate states. The two-pion cut is indicated by the  (red) dotted line. The vertex
	$F_V P A^\eta_{\pi\pi\gamma}$ indicates the $\eta\to \pi^+\pi^-\gamma$ transition form factor, while
	the other vertex corresponds to the two-pion vector form factor $F_V^\ast$ times the electric charge $\e$, see 
	Eq.~(\ref{eq:xDisc1}) for more details.}
 	\label{fig:etatogammagamma}
\end{figure}

The discontinuity of the isovector part of the $\eta\to \gamma\gamma^\ast$ decay amplitude
for $Q^2<(4m_\pi)^2$
is driven by the on-shell two-pion intermediate states, 
see Figure~\ref{fig:etatogammagamma}.
%%%%%%%%%%%%%%%%%%%%%%%%%%%%%%%%%%%%%%%%%%%%%%%%%%%
%%%%%%%%%%%%%%%%%%%%%%%%%%%%%%%%%%%%%%%%%%%%%%%%%%%%
Especially, one finds 
\begin{eqnarray}
 \lefteqn{\text{Disc} \ \mathcal{A}_{1}^{\rho\mu}} \nonumber\\
 &=& i (2\pi)^4 \int d \Phi_2 \,
          \mathcal M^\mu \!\left( \eta (p_\eta) \to \pi^+ (p_1) \pi^- (p_2) \gamma (p_\gamma) \right) \nonumber \\
        &&  \mbox{} \times {{\mathcal M}^{\rho}}^\ast \left( \pi^+ (p_1) \pi^- (p_2) \to \gamma^\star (p_\gamma) \right) 
                                                 \nonumber \\
  &=& i (2\pi)^4 \int d \Phi_2\, P(Q^2) F_V(Q^2) A_{\pi\pi\gamma}^\eta \, \epsilon^{\mu\nu\alpha\beta}
  (p_\gamma)_\nu (p_1)_\alpha (p_2)_\beta\nonumber \\
                    &&       \mbox{}       \times {\e}  F_V(Q^2)^\ast  (p_1 - p_2)^\rho \nonumber \\
                                  &=& i  (2\pi)^4\,{\e}A_{\pi\pi\gamma}^\eta\, P(Q^2)\left |F_V(Q^2)\right|^2  \epsilon^{\mu\nu\alpha\beta} (p_\gamma)_\nu \nonumber \\
&& \mbox{} \times                                 \int d \Phi_2 \, (p_1 - p_2)^\rho (p_1)_\alpha (p_2)_\beta\,, \label{eq:xDisc1}
\end{eqnarray}
where $\e$  is the unit of electric charge.
Defining  $k \equiv (p_1{-}p_2)/2$
and  $Q\equiv p_1+p_2$ we get 
\begin{eqnarray}
\text{Disc} \  \mathcal A_{1}^{\rho \mu} 
                                   &=& 2i  (2\pi)^4 \,{\e}\,
                                   A_{\pi\pi\gamma}^\eta\, P(Q^2) \left|F_V(Q^2)\right|^2 
                                   \epsilon^{\mu\nu\alpha\beta}  
                                   (p_\gamma)_\nu 
                                   \nonumber \\
                                   && \mbox{}\times \int d \Phi_2 \ k^\rho k_\alpha Q_\beta\,. 
\end{eqnarray}
In the $\eta$ rest frame 
we have
$\vec Q=-\vec p_\gamma$ and therefore
\[
 \epsilon^{\mu\nu\alpha\beta} ({p_\gamma})_\nu Q_\beta =\sqrt{s_\eta} \epsilon^{mab}Q^b 
\]
 where $\epsilon_{0123}=-\epsilon^{0123}= +1$
and  $m, \ a, \ b$ denote the spatial components for the Lorentz indices $\mu,
\ \alpha, \ \beta$, respectively.
We thus  get,
using 
\begin{eqnarray*}
(2\pi)^4 d\Phi_2 k^r k^a &=& d\Omega \frac1{32\pi^2}\sigma_\pi(Q^2) k^rk^a\\ 
&=& \frac1{32\pi^2}\left(\frac{4\pi}{3}\right)\sigma_\pi(Q^2) \vec k\, ^2\delta^{ra} 
\end{eqnarray*}
and
$\vec k\, ^2= (Q^2-4m_\pi^2)/4 = {Q^2}\sigma_\pi^2(Q^2)/4$,
\begin{eqnarray}
\text{Disc} \ \mathcal A_{1}^{rm} &=& 2i\,\pi\, \e  
A_{\pi\pi\gamma}^\eta \sqrt{s_\eta}\, \epsilon^{mrb} {p_\gamma}^b 
 \nonumber \\
&& \mbox{} \times 
   \frac{Q^2}{96 \pi^2} \sigma_\pi(Q^2)^3\, P(Q^2) \left| F_V(Q^2)\right|^2\ .
\end{eqnarray}
Due to $ \text{Disc}\,  \mathcal A_{1}^{rm} =
 2 i \  {\rm Im} \ \mathcal A_{1}^{rm} $, 
we may then write a once-subtracted dispersion integral 
for  
$\Delta\mathcal A_{1}^{rm} (Q^2)$ intro\-du\-ced in Eq.~(\ref{Adecomp}):
\begin{eqnarray}
  \lefteqn{\Delta\mathcal A_{1}^{rm} (Q^2)
   =  {\e} A_{\pi\pi\gamma}^\eta\sqrt{s_\eta}\epsilon^{mrb} p_\gamma^b } \nonumber \\
&&  
    \qquad\quad\mbox{}\times \frac{Q^2}{96 \pi^2} \int_{4 m_\pi^2}^\infty 
                                                ds^\prime \sigma_\pi(s')^3
                                                \ P(s^\prime)
                                                \ \frac{|F_V(s^\prime)|^2}{s^\prime
                                                  - Q^2 - i\epsilon} \, ,
   \label{adispsub}
\end{eqnarray}
where the subtraction constant will be absorbed in  the double-on-shell amplitude $\mathcal A^{rm}(0)$.
The  $\eta\to \gamma\gamma^\star$ transition form factor is
defined via, {\it cf.} Eq.\,(\ref{Armzero}),
\begin{equation}
 \mathcal A^{rm} (Q^2) = {A_{\gamma\gamma}^\eta}\,
 \sqrt{s_\eta}\,\epsilon^{mrb} p_\gamma^b\ 
 F_{\eta\gamma^\ast\gamma}(Q^2,0)
 \label{Fdef}
\end{equation}
as
\begin{eqnarray}
\lefteqn{F_{\eta\gamma^\star\gamma}(Q^2,0) \equiv 1 
 +  \Delta F_{\eta\gamma^\star\gamma}^{(I=1)}(Q^2,0) 
 +  \Delta F_{\eta\gamma^\star\gamma}^{(I=0)}(Q^2,0) } \nonumber\\ 
&=& 1 + \kappa_\eta
\left(\frac{Q^2}{96
   \pi^2 f_\pi^2}\right)    \int_{4 m_\pi^2}^\infty 
                                                ds^\prime\, \sigma_\pi(s')^3
                                                \ P(s^\prime)
                                                \ \frac{|F_V(s^\prime)|^2}{s^\prime
                                                  - Q^2 - i\epsilon} \nonumber \\ 
                                                  & & \mbox{} \ \ + \Delta F_{\eta\gamma^\star\gamma}^{(I=0)}(Q^2,0) \ ,
 \label{eq:DR}
\end{eqnarray}
where the isovector contribution $\Delta F_{\eta\gamma^\star\gamma}^{(I=1)}(Q^2,0)$ is specified in
the second line and where the
isoscalar one is defined to vanish in the on-shell limit as well,  {\it i.e.}
$\Delta F_{\eta\gamma^\star\gamma}^{(I=0)}(0,0)=0$. Furthermore, we adopt the prefactor 
$\kappa_\eta\equiv  {{\e}A_{\pi\pi\gamma}^\eta {f_\pi^2}}/ A_{\gamma\gamma}^\eta$, 
with $f_\pi =92.2\,\text{MeV}$  the pion decay constant~\cite{pdg:2012}, introduced here
for later convenience. Note, in the SU(3) chiral limit one has $\kappa_\eta = 1$.
Based on Eq.~(\ref{agamgamdef}) and Eq.~(\ref{eq:main}) $\kappa_\eta$ can be fixed directly from data.

The pertinent slope parameters are defined via
\begin{equation}
 F_{\eta\gamma^\star\gamma}(Q^2,0)=
  1 + \left( b_\eta^{(I=1)}+b_\eta^{(I=0)}\right) Q^2
 +{\mathcal O}(Q^4)  \ .
\end{equation}
Thus, from Eq.~(\ref{eq:DR}) we get the following integral representation for
the isovector component of the slope parameter
\begin{equation}
 b_\eta^{(I=1)} =
\frac{\kappa_\eta}{6(4\pi f_\pi)^2} \int_{4 m_\pi^2}^\infty 
                                                \frac{ds^\prime}{s^\prime} \ \sigma_\pi(s')^3
                                                \ P(s^\prime)\left|F_V(s^\prime)\right|^2 \ .
  \label{eq:slope}
\end{equation}
The isovector part of the form factor is
model-independent, since it can be expressed fully in terms of experimental
observables. Those are the branching ratios  (or partial decay widths) of $\eta\to \pi^+\pi^-\gamma$ and
$\eta\to \gamma \gamma$, to fix the prefactor $\kappa_\eta$, the slope parameter $\alpha$
from the spectral shape of $\eta/\eta'\to \pi^+\pi^-\gamma$
({\it cf.\/} Ref.~\cite{Stollenwerk} and Eq.~(\ref{alphaval})) and the pion vector form factor.
As will be demonstrated below, the uncertainty from our ignorance about the high-$Q^2$
behavior of both $P(Q^2)$ as well as $F_V(Q^2)$ can be estimated reliably.
The isoscalar component of the slope parameter, $b_\eta^{(I=0)}$, will be discussed in the
next section.

%%%%%%%%%%%%%%%%%%%%%%%%%%%%%%%%%%%%%%%%%%%%%%%%%%
%%%%%%%%%%%%%%%%%%%%%%%%%%%%%%%%%%%%%%%%%%%%%%%%%%

\section{Model for the  isoscalar contribution of the slope parameter}
\label{sec:model}

The two-pion contribution is almost purely isovector (up to a small contribution from
the $\omega$ contributing via $\rho$-$\omega$ mixing). However, the full slope parameter 
contains also an isoscalar contribution. To quantify this part,
it is necessary to construct a model. Especially we will assume that, in the spirit of vector
meson dominance (VMD),
the isoscalar part is saturated by the contribution of two lowest isoscalar vector-meson 
resonances, $\omega$ and $\phi$ which are both narrow.
 However, as we will demonstrate, the model parameters are largely constrained by data and,
at least in case of the $\eta$, 
the total isoscalar contribution is small.

We  chose as a model ansatz for
the isoscalar contribution
to the transition form factor of the $\eta$ 
\begin{equation}
 \Delta F_{\eta \gamma^\star \gamma}^{(I=0)}(Q^2,0) 
 =  \frac{w_{\eta\omega\gamma} \,Q^2}{m_\omega^2 - Q^2 -i m_\omega\Gamma_\omega} 
    +\frac{w_{\eta\phi\gamma} \,Q^2}{m_\phi^2 - Q^2 -i m_\phi \Gamma_\phi }  \ .
 \label{eq:TFF-Delta-isoscal}
\end{equation} 
Here 
$m_\omega$ ($\Gamma_\omega$) and  $m_\phi$ ($\Gamma_\phi$) denote the mass (total width) 
of the $\omega$ and
$\phi$ meson, respectively, as given in Ref.~\cite{pdg:2012}.
In order to determine the weight factors $w_{\eta \omega\gamma}$ and  $w_{\eta \phi\gamma}$, 
we now follow two paths: (i)
 we employ the VMD model of Ref.~\cite{Landsberg}
 to determine the {\em magnitude and sign} of the weight factors;   (ii) we fix
the modulus of the weight factors
from data directly, however, we still need to stick to the phases
as given in Ref.~\cite{Landsberg}.

In the VMD model of Ref.~\cite{Landsberg} one finds\footnote{Clearly, in that work also an expression
for the isovector contribution is given, however, we will omit this part here since we fix it model-independently
from dispersion theory.}
\begin{equation}
   w_{\eta\omega\gamma} = \frac{\frac{1}{9}}{1+\frac{1}{9}-\frac{\sqrt{2}}{3}\beta_\eta} = \frac{1}{8}\ , \ \ 
  w_{\eta\phi\gamma}=\frac{-\frac{\sqrt{2}}{3}\beta_\eta}{1+\frac{1}{9}-\frac{\sqrt{2}}{3}\beta_\eta} = -\frac{2}{8}
 \label{eq:weight-standard}
\end{equation}  
in terms of a  one-angle $\eta$-$\eta'$ mixing scheme 
\begin{equation}
             \beta_\eta   =  \frac{2}{3} \left[ \frac{\sqrt{2}\cos\theta_P + \sin\theta_P}{\cos\theta_P - \sqrt{2}\sin\theta_P} \right] 
                              =\frac{\sqrt{2}}{3}  \approx 0.47 \ .        
\end{equation} 
Here   we applied                          
the standard value in chiral perturbation theory (ChPT), 
\begin{equation}
  \theta_P  =  \arcsin(-1/3)  \approx -19.5^\circ\,,
 \label{eq:standard-mixing}
 \end{equation}
 see, {\it e.g.\/}, Ref.~\cite{Bijnens_1990},
for the mixing angle $\theta_P$ of the pseudoscalar nonet. This value is consistent with both a one-loop analysis
for the mass matrix and the two-photon decays of $\eta$ and $\eta'$~\cite{barry}.

 The resulting expression for the isocalar
contribution to the  slope of the $\eta$ transition
form factor is then given by
\begin{equation}
         b^{(I=0)}_\eta  
             =  \frac{w_{\eta \omega\gamma}}{m_\omega^{2}}+ \frac{w_{\eta\phi\gamma}}{m_\phi^{2}} 
                 \approx    - 0.036\,\text{GeV}^{-2} \ .
                 \label{eq:b-zero}
\end{equation}
The isoscalar component   (\ref{eq:b-zero}) turns out to be smaller than the uncertainty of
our full calculation, when  the standard value for the  mixing
angle,  $\theta_P  =\arcsin(-1/3)$, is used. 
In case of the $\eta'$, however, this mixing angle  
leads to the weights
\begin{equation}
   w_{\eta\omega\gamma} = \frac{\frac{1}{9}}{1+\frac{1}{9}-\frac{\sqrt{2}}{3}\beta_{\eta'}} = \frac{1}{14}\ , \ \ 
  w_{\eta\phi\gamma}=\frac{-\frac{\sqrt{2}}{3}\beta_{\eta'}}{1+\frac{1}{9}-\frac{\sqrt{2}}{3}\beta_{\eta'}} = \frac{4}{14},
 \label{eq:weight-standard-etap}
\end{equation}  
since
$-\beta_\eta \sqrt{2}/3=- 2/9$ in Eq.~(\ref{eq:weight-standard}) has to be replaced by $+\beta_{\eta'}\sqrt{2}/3 = +4/9$ with $\beta_{\eta'} = 4/(9 \beta_{\eta}) = 2\sqrt{2}/3$. 
This results in
a positive and comparably large shift of $0.39\,\text{GeV}^{-2}$ for
$b_{\eta'}^{(I=0)}$.

Based on an analysis of a large set of data, Refs.~\cite{Benayoun_EPJC55,Benayoun_EPJC65} report a mixing angle of about $-10.5^\circ$ (see also \cite{Benayoun_EPJC17b,Benayoun_EPJC72}). However,
within that approach other parameters change as well and, based on this model class, one gets, respectively, 
$b_\eta^{(I=0)}=-0.023\,\text{GeV}^{-2}$ and $b_{\eta'}^{(I=0)}= 0.30\,\text{GeV}^{-2}$---rather close to the values given above.
The spread between the two different results for the isoscalar contributions will be included in the uncertainties. 
If, on the other hand, we  had used an angle of $-10.5^\circ$ directly
in Eq.~(\ref{eq:b-zero}), the isoscalar correction to $b_\eta$ would have been as large as $-0.15\,\text{GeV}^{-2}$
while that to $b_{\eta'}$ would have been $0.34$ GeV$^{-2}$.

So far we fully relied on the VMD model to fix the contributions
from the two isoscalar resonances to the transition form factor. However, empirical input 
from Ref.\,\cite{pdg:2012} may be used
to determine the moduli of the weight factors $w_{\eta\omega\gamma}$ and $w_{\eta\phi\gamma}$
in the ansatz (\ref{eq:TFF-Delta-isoscal})\footnote{\label{foot:sign}In principle even the sign of the weights,
which are assumed to be real-valued, can also be inferred from
the $e^+ e^- \to \gamma\eta$ data, namely from the asymmetric behavior of the cross section slightly
below and slightly above the resonance pole(s)---for a comparison with 
data see, {\it e.g.}, \cite{CMD2-epem,SND-epem} 
and references therein.}.
For this one matches
the relativistic version of the Breit-Wigner cross section at the narrow iso\-scalar vector meson  pole(s) 
\begin{equation}
  \sigma(e^+ e^- {\to} \eta\gamma)|_{s=m_V^2} = \frac{12\pi\, \text{BR}(V\to \eta\gamma) \,\text{BR}(V\to e^+ e^-)}{m_V^2} \, ,
  \label{eq:sig_epem2etaga_BR}
 \end{equation} 
$V = \omega, \phi$ (see {\it e.g.} \cite{VEPP-2M}),
with 
\begin{equation}
 \sigma(e^+ e^- \to \eta \gamma) = \frac{2}{3} \e^2 \Gamma^\eta_{\gamma\gamma}
 \left(\frac{s-m_\eta^2}{s m_\eta}\right)^3 \left| \Delta F_{\eta \gamma^\star \gamma}^{(I=0)}(s,0) \right|^2  \,
  \label{eq:sig_epem2etaga_FF}
\end{equation}
evaluated at $s=m_V^2$,
{\it cf.} Ref.~\cite{krakau}\footnote{Note, however, that in this reference 
the factor $2/3$ on the right-hand side is missing---compare, {\it e.g.}, with the correct expression of 
\cite{Benayoun_EPJC65}.}.  Since the resonances are narrow, the contribution
from the isoscalar part of the constant term, $F_{\eta \gamma^* \gamma}^{(I=0)}(0,0)$, can be neglected at the vector meson poles.
Inserting
the branching ratios (BR) for the decays $\omega\to \eta\gamma$ and $\omega\to e^+e^-$, which are
tabulated in Ref.\,\cite{pdg:2012}, we get
$w_{\eta\omega\gamma} \approx (0.78\pm 0.04){\times}1/8$, 
while the branching ratios for the decays $\phi\to \eta\gamma$ and $\phi\to e^+e^-$
give the result
$w_{\eta\phi\gamma} \approx (0.75\pm 0.03){\times}(-2/8) $.

Thus the fit to data reduces
the weights for the standard-mixing-angle case  
approximately  by a factor $3/4$, such that the isoscalar contribution to the slope of
the $\eta$ transition form factor reads
$b_\eta^{(I=0)} \approx -0.022\,\text{GeV}^{-2}$, which is almost the
result of the mixing scheme of Refs.\,\cite{Benayoun_EPJC55,Benayoun_EPJC65} and about 60\%
of the result (\ref{eq:b-zero}) of the standard mixing case (\ref{eq:standard-mixing}). This deviation
 is included in the final uncertainty. 
 
In 
case of the $\eta'$, the above steps can 
be copied for the cross section $\sigma(e^+e^-\to \eta'\gamma)$ at 
the $\phi$ pole. The corresponding weight is then
$w_{\eta' \phi \gamma} \approx (0.54 \pm 0.02)\times 4/14$, {\it i.e.} slightly bigger than half of the weight for the
standard mixing-angle scenario. However, additional theoretical input is  needed to determine
the weight $w_{\eta'\omega\gamma}$, since the decay $\omega\to\eta'\gamma$ is of course kinematically forbidden.
For that purpose we rewrite, always at a specified  $V$ pole and with $P=\eta,\eta'$, respectively, Eqs.~(\ref{eq:sig_epem2etaga_BR}) 
and    (\ref{eq:sig_epem2etaga_FF}) with input of (\ref{eq:TFF-Delta-isoscal}) as
\begin{equation}
  w^2_{P V\gamma} 
         = \frac{12\pi\, g_{V\to P\gamma}^2 \, \Gamma_{V\to e^+ e^-} / m_V^3}
       {16 \e^2 \,\Gamma^P_{\gamma\gamma}/m_P^3 }
  = \frac{12\pi\, g_{V\to P\gamma}^2 \, \Gamma_{V\to e^+ e^-}}
       {\alpha_{\text{em}} |\mathcal A_{P\to \gamma\gamma}|^2 m_V^3} \,.
 \label{eq:ws_P2Vga}
\end{equation}
Here
$\alpha_{\text{em}}=\e^2/(4\pi)$ is the electromagnetic fine structure constant. Furthermore,
the standard  $p$-wave expression for the $V\to P\gamma$ decay width, 
\begin{equation}
 \Gamma_{V\to P\gamma} = \frac{g_{V\to P \gamma}^2}{3 m_V^2} \left(\frac{m_V^2 - m_P^2}{2 m_V} \right)^3\,,
  \label{eq:Ga_V2Pga}
\end{equation}
and the $P$-analog of  Eq.~(\ref{agamgamdef}) 
have been inserted.
Equation (\ref{eq:ws_P2Vga})  
holds of course for all three cases that we have
discussed above, $w_{\eta \omega\gamma}$, $w_{\eta \phi\gamma}$ and $w_{\eta'\phi\gamma}$. Now, in the
remaining  $w_{\eta'\omega\gamma}$ case we use in addition the usual $p$-wave formula for the decay $P\to V\gamma$,
\begin{equation}
 \Gamma_{P \to V\gamma} = \frac{g_{P\to V \gamma}^2}{m_P^2} \left(\frac{m_P^2 - m_V^2}{2 m_P} \right)^3\,,
  \label{eq:Ga_P2Vga}
\end{equation}
with the theoretical understanding 
that the square of the dimensional coupling constants satisfy
$g_{P\to V \gamma}^2 = g_{V\to P \gamma}^2$.
 Then again the  branching ratios or partial decay widths 
tabulated in Ref.\,\cite{pdg:2012}
are sufficient to determine $w_{\eta'\omega\gamma}$
in magnitude---the sign follows from Eq.~(\ref{eq:weight-standard-etap}).
The final result is $w_{\eta'\omega\gamma}= (1.27 \pm 0.07)\times 1/14$,
which is approximately 30\% bigger than the one of the standard-mixing scenario. In summary, the slope at the origin of the
$\eta'$ transition form factor reads $b_{\eta'}^{(I=0)}\approx 0.30\,\text{GeV}^{-2}$, which is compatible with the result of 
the mixing scheme of Refs.\,\cite{Benayoun_EPJC55,Benayoun_EPJC65}  and about 75\% of the result of the
standard-mixing scenario. 

In order to give a conservative estimate of this contribution, 
we take for its central value  the arithmetic mean of the two results reported above, while the
difference determines the uncertainty range: $b_{\eta'}^{(I=0)} = (0.34\pm 0.05)\,\text{GeV}^{-2}$.
Compared to this uncertainty the uncertainties from the weight factors $w_{PV\gamma}$
turn out to be negligible, when added in quadrature.

Note that neither the model-independent isovector part in (\ref{eq:DR}) nor the additional isoscalar contributions
(\ref{eq:TFF-Delta-isoscal}) to the $\eta$ transition form factor  vanish in the limit $Q^2\to \infty$. This fact
is closely tied to the  choice of the {\em once-subtracted} form of the dispersion integral in Sect.~\ref{sec:dispersion} 
that has the inherent property that 
the subtraction constant must be determined by empirical input.  In fact, we 
rather prefer to determine the transition form factor from the correct
low-energy empirical input than to rely on a loose extrapolation to perturbative QCD which favours the
vanishing of the transition form factor at  $Q^2\to \infty$~\cite{BL,Brodsky_2011,Bijnens_Pers}.

%%%%%%%%%%%%%%%%%%%%%%%%%%%%%%%%%%%%%%%%%%%%%%%%%%
%%%%%%%%%%%%%%%%%%%%%%%%%%%%%%%%%%%%%%%%%%%%%%%%%%

\section{Results}
\label{sec:results}

The uncertainties for the evaluation of the isovector part of the transition form factor
 emerge from those of the experimental branching fractions (collected in the
prefactor $\kappa_\eta$) and from the value of $\alpha$ ({\it cf.\/} Eq.~(\ref{alphaval})). 

Formally the integral of Eq.~(\ref{eq:DR}) runs up to infinity. On the other hand we
can control its input, especially $P(Q^2)$, only in the regime up to $Q^2=1\,\text{GeV}^2$.
In order to demonstrate that the relevant contributions indeed come from the
regime below $1\ \text{GeV}^2$,  we follow Refs.~\cite{Schneider,Bastian_Martin} and investigate the 
{\em un}-subtracted dispersion integral, 
analog to Eq.~(\ref{adispsub}),
which provides a sum rule for $A^\eta_{\gamma\gamma}$. Namely the iso\-vector part 
of the $\eta\to \gamma\gamma$ amplitude should satisfy
\begin{equation}
A_{\gamma\gamma}^{\eta \,(I=1)}= {\e}\, A_{\pi\pi\gamma}^\eta 
 \ \frac{1}{96 \pi^2} \int_{4 m_\pi^2}^\infty 
                                                ds^\prime \sigma_\pi(s')^3
                                                \ P(s^\prime)
                                                \ |F_V(s^\prime)|^2  .
                                                \label{check_isovec}
\end{equation}
To estimate the {\em model-dependent} isoscalar contribution to the form
factor normalization, we need to replace in the numerators of Eq.~(\ref{eq:TFF-Delta-isoscal}) 
the factors $Q^2$ by the corresponding $m_V^2$. Using Eq.~(\ref{Fdef}) this gives, 
\begin{equation}
A_{\gamma\gamma}^{\eta \,(I=0)}=\left( w_{\eta\omega\gamma} + w_{\eta\phi\gamma} \right)A^\eta_{\gamma\gamma} \ .
\end{equation}
With this we get 
\begin{eqnarray}\nonumber
A^\eta_{\gamma\gamma} &=& A_{\gamma\gamma}^{\eta \,(I=1)} + A_{\gamma\gamma}^{\eta \,(I=0)} \\
&=& A_{\gamma\gamma}^{\eta \,(I=1)} +(w_{\eta\omega\gamma} + w_{\eta\phi\gamma})A^\eta_{\gamma\gamma}   \ .
                                                \label{check_full}
\end{eqnarray}
If $P(s)$ were linear up to infinite energies, the 
integral in (\ref{check_isovec}) would be  formally log-divergent, since $F_V(s)\sim 1/s$ for large
values of $s$. 
However, the goal here is to confirm that all relevant physics
is located 
below $1\,\text{GeV}^2$.  And indeed, if the pertinent integral in  Eq.~(\ref{check_isovec})
is truncated at $1\,\text{GeV}^2$, the right-hand side of the sum rule
(\ref{check_full}) overestimates the left-hand one by only $(7\pm 5)\%$.
If we vary the upper integration range between $s = m_{\eta'}^2$
and $s=1.15$ GeV$^2$ (the largest value of $s$
where the form factor shown in the upper panel of Fig.~\ref{fig:overOmnes}
is still linear), 
the mismatch between  the right-hand and left-hand side  increases to  $(9 \pm 11)\%$.
This provides strong evidence that the once-subtracted integral of Eq.~(\ref{eq:slope}) and thus also
of Eq.~(\ref{eq:DR}) should provide reliable results, when being cut at  or slightly below $1\,\text{GeV}^2$.
For the $\eta$ decay the isoscalar contribution turns out to be negligible. 

To get a conservative estimate for the possible impact of higher values of $s$
in the integral of Eq.~(\ref{eq:DR}) and Eq.~(\ref{eq:slope}), respectively, we  also evaluated the integral using $s_{\rm max}= 2\,\text{GeV}^2$ ---an
increase to  $s_{\text{max}}= 3\,\text{GeV}^3$ did not alter the displayed results. For that purpose we
continue $P(s)$ linearly in combination with the  two form factors $F_V(Q^2)_{e^+e^-}$,
and  $F_V(Q^2)_{\tau}$ 
introduced at the end of Sect.~\ref{sec:eta2pipigam}. This procedure lead to some increase in 
the transition form factor,  the largest results were obtained with the maximum
input value for $\alpha$ from Eq.~(\ref{alphaval}) in combination with the $\tau$ form factor,  
$F_V(Q^2)_{\tau}$.

%%%%%%%%%%%%%%%%%%%%%%%%%%%%%%%%%%%%%%%%%%%%%%%%%%%%
\begin{figure} % Figure 3
\begin{minipage}{\columnwidth}
\centering
   \includegraphics[width=\textwidth]{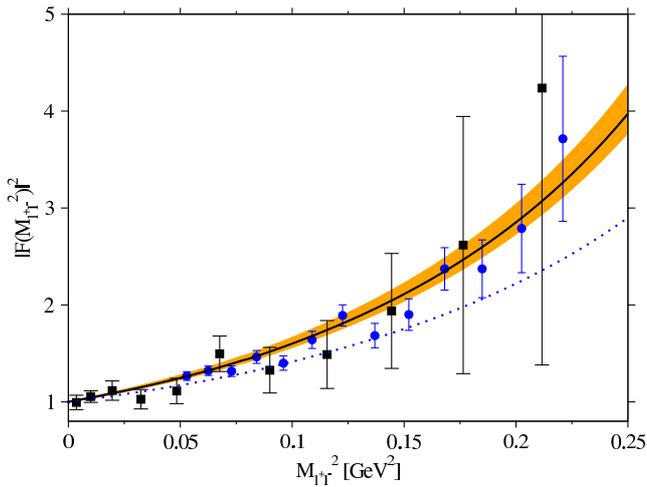}
 \end{minipage}  
   \caption{The squared modulus of the $\eta\to\gamma\gamma^*$ transition form factor as function
   of the invariant mass square, $M_{l^+l^-}^2$, of the (electron or muon) dilepton pair from the subsequent decay 
   $\gamma^* \to l^+ l^-$. The
   results of Eq.~(\ref{eq:DR}) with input from Eqs.~(\ref{eq:TFF-Delta-isoscal}) and 
   (\ref{eq:weight-standard}) are
   compared with the two most recent measurements from Refs.~\cite{NA60_2011,CB_TAPS_2011}, which are 
   displayed as solid dots and squares, respectively.
   The (orange) band shows the spread of our results emerging from the uncertainty
   in $\alpha$ (deduced from a fit to $\eta\to\pi\pi\gamma$ of Ref.~\cite{KLOE_2013}---{\it cf. \/} Eq.~(\ref{alphaval})),
   from the variation of the end point $s_\text{max}$ of the integral (from $m_{\eta'}^2$ to $2\,\text{GeV}^2$), the
   applied
   form factors and the uncertainties of branching ratios  entering the prefactor.   
       Solid line: our central result (with $\alpha =1.32\ \text{GeV}^{-2}$, and 
      $s_{\rm max}=1\ \text{GeV}^2$).
        Dotted line:  dispersion integral with $\alpha=0$  and $s_{\rm max}=1\ \text{GeV}^2$.\label{fig:formfactor}}
\end{figure}
%%%%%%%%%%%%%%%%%%%%%%%%%%%%%%%%%%%%%%%%%%%%%%%%%

The resulting spread for the $\eta\to\gamma\gamma^*$ transition form factor that emerges from the calculation,
including the uncertainties mentioned above and with the upper limit of integration varied from $s_{\rm max}=m_{\eta'}^2$ to
$2\,\text{GeV}^2$, 
is shown as the (orange) band in Figure~\ref{fig:formfactor}.

The formalism allows one to disentangle effects from the $\pi\pi$-interactions, which are universal, from those 
of the decay vertex, which are reaction specific. Thus it is interesting to investigate how much of the form factor
emerges from the two-pion interactions and how much from the production vertex. We therefore
show as the (blue) dotted line in Figure~\ref{fig:formfactor} the result for $\alpha=0$. Thus about 20\%
of the slope of the $\eta$ transition form factor results from the decay vertex while 80\% come from the
$\pi\pi$ interactions.

%%%%%%%%%%%%%%%%%%%%%%%%%%%%%%%%%%%%%%%%%%%%%%%%%%
\begin{table*}    % Table 1
\caption{Comparison of our result for the slope parameter $b_\eta$ as given in
Eq.~(\ref{res_b_full})
with experimental as well as previous theoretical investigations.
 The results for the theoretical works (except \cite{Escribano}) are taken from
Table~II of Ref.~\cite{Ametller_1992}. The experimental result $b_\eta = (1.6\pm 2.0)\,\text{GeV}^{-2}$ of  
Ref.\,\cite{SND_2001}  (for the process $\eta \to e^+ e^- \gamma$) is not included because of its large uncertainty.\label{tab:results} }
\centering
\begin{tabular}{clclc}
\hline\noalign{\smallskip}
Type &Process & Ref. & $b_\eta \,[\text{GeV}^{-2}]$  & \parbox{6cm}{\includegraphics[width=6cm]{label_up}}\\  
\noalign{\smallskip}
Exp. &$\eta\to \mu^+\mu^-\gamma$ & \cite{Lepton_G_1980} &  $1.90 \pm 0.40$ &\parbox{6cm}{\includegraphics[width=6cm]{tab1a}} \\
Exp. &$e^+e^-\to e^+ e^- \gamma\gamma^*\to e^+e^-\eta$ & \cite{TPC2gamma}  & {$2.04 \pm 0.47$} & \parbox{6cm}{\includegraphics[width=6cm]{tab1b}} \\
Exp.& $e^+e^-\to e^+ e^- \gamma\gamma^*\to e^+e^-\eta$ &\cite{Cello_1991} &  {$1.42 \pm 0.20$}  & \parbox{6cm}{\includegraphics[width=6cm]{tab1c}}\\
Exp.& $\eta\to \mu^+\mu^-\gamma$ & \cite{NA60_2009}  & $1.95 \pm 0.17$ &  \parbox{6cm}{\includegraphics[width=6cm]{tab1d}} \\
Exp. & $\eta\to \mu^+\mu^-\gamma$ & \cite{NA60_2011}  &$1.95 \pm 0.07$ &  \parbox{6cm}{\includegraphics[width=6cm]{tab1e}} \\
Exp. & $\eta\to e^+e^-\gamma$ & \cite{CB_TAPS_2011}   &$1.92 \pm 0.37$ &  \parbox{6cm}{\includegraphics[width=6cm]{tab1f}}\\
Theory   &VMD  & \cite{QL1,QL2,QL3}       & $1.78$ &  \parbox{6cm}{\includegraphics[width=6cm]{tab1g}} \\
Theory & Quark loop  & \cite{QL1,QL2,QL3}    &$1.69$ &  \parbox{6cm}{\includegraphics[width=6cm]{tab1h}} \\
Theory & Brodsky-Lepage & \cite{BL}          &$1.21$ &  \parbox{6cm}{\includegraphics[width=6cm]{tab1i}} \\
Theory &1-loop ChPT   &\cite{Ametller_1992}                &$1.69$ &  \parbox{6cm}{\includegraphics[width=6cm]{tab1j}} 
\\
Theory & Pad\'e  approx.\ fit to \cite{Cello_1991,CLEO_1998,BaBar_2011} data & \cite{Escribano} & $1.99\pm 0.16 \pm 0.11$ & \parbox{6cm}{\includegraphics[width=6cm]{tab1l}} \\
Theory & Dispersion integral &This work                                                  &$2.05\ \mbox{}^{+0.22}_{-0.10}$ 
&  \parbox{6cm}{\includegraphics[width=6cm]{tab1k}} \\
             &                                &                               &    & \parbox{6cm}{\includegraphics[width=6cm]{label}} \\
\noalign{\smallskip}\hline
\end{tabular}
\end{table*}
%%%%%%%%%%%%%%%%%%%%%%%%%%%%%%%%%%%%%%%%%%%%%%%%%%

%%%%%%%%%%%%%%%%%%%%%%%%%%%%%%%%%%%%%%%%%%%%%%%%%%
\begin{table*}   % Table 2
\caption{Comparison of our result  for the slope parameter $b_{\eta'}$ 
as given in 
Eq.~(\ref{res_bp_full})
with experimental as well as previous theoretical investigations, under the additional assumption that the 
parameter $\alpha$, {\it cf.\/} Eq.~(\ref{alphadef}), is the same for both $\eta$ and $\eta'$ decays.
The results for the various experimental and theoretical works  (except \cite{Escribano}) are taken from
Ref.~\cite{Ametller_1992}.  \label{tab:results_prime} }
\centering
\begin{tabular}{clclc}
\hline\noalign{\smallskip}
Type &Process & Ref. & $b_{\eta'}\, [\text{GeV}^{-2}]$ & 
\parbox{6cm}{\includegraphics[width=6cm]{etap_label_up}} \\
\noalign{\smallskip}
Exp. &$\eta'\to \mu^+\mu^-\gamma$ & \cite{Lepton_G_1979,Lepton_G_1980}&$1.69\pm 0.79$ & \parbox{6cm}{\includegraphics[width=6cm]{etap_exp1}} \\
Exp. &$e^+e^-\to e^+ e^- \gamma\gamma^*\to e^+e^-\eta'$ & \cite{TPC2gamma}  &$1.38\pm 0.23$& \parbox{6cm}{\includegraphics[width=6cm]{etap_exp2}} \\
Exp.& $e^+e^-\to e^+ e^- \gamma\gamma^*\to e^+e^-\eta'$ & \cite{Cello_1991}  & $1.60\pm 0.16$ & \parbox{6cm}{\includegraphics[width=6cm]{etap_exp3}}\\
Theory &VMD     & \cite{QL1,QL2,QL3} &$1.45$ & \parbox{6cm}{\includegraphics[width=6cm]{etap_th1}} \\
Theory &Quark loop   & \cite{QL1,QL2,QL3} & $1.42$ & \parbox{6cm}{\includegraphics[width=6cm]{etap_th2}} \\
Theory & Brodsky-Lepage     & \cite{BL}      &   $2.30$  & \parbox{6cm}{\includegraphics[width=6cm]{etap_th3}} \\
Theory &1-loop ChPT        & \cite{Ametller_1992} &$1.60$ & \parbox{6cm}{\includegraphics[width=6cm]{etap_th4}} \\
Theory & Pad\'e  approx.\ fit to \cite{Cello_1991,CLEO_1998,BaBar_2011} data & \cite{Escribano} & $1.49\pm 0.17 \pm 0.09$ & \parbox{6cm}{\includegraphics[width=6cm]{etap_th5}} \\
Theory & Dispersion integral  &This work                                           &  $1.53\  \mbox{}^{+0.15}_{-0.08}$ & \parbox{6cm}{\includegraphics[width=6cm]{etap_thiswork_new}} \\
             &                                &                          &         & \parbox{6cm}{\includegraphics[width=6cm]{etap_label}} \\
\noalign{\smallskip}\hline
\end{tabular}
\end{table*}
%%%%%%%%%%%%%%%%%%%%%%%%%%%%%%%%%%%%%%%%%%%%%%%%%%

The isovector contribution   of the slope of the transition amplitude
is determined to be
\begin{equation}
 b_\eta^{(I=1)} = \left(2.09\  \mbox{}^{+0.21}_{-0.11} \right)\text{GeV}^{-2}  \, .
   \label{res_b_isovector}
\end{equation}
The uncertainties include those of the branching ratios, the parameter 
$\alpha$, the form factor as well as the range of integration.
If the isoscalar contribution is added to $b_\eta^{(I=1)}$,
we get for the full slope of the transition form factor
\begin{equation}
   b_\eta = \left(2.05 \ \mbox{}^{+0.22}_{-0.10}  \right)\text{GeV}^{-2} 
 \label{res_b_full}
\end{equation}
for the standard value for the $\eta$-$\eta'$ mixing angle $\theta_P = - 19.5^\circ$.
The uncertainties are analogous to those shown in Eq.~(\ref{res_b_isovector}).
The result (\ref{res_b_full}) 
is compatible with all recent 
experimental results, but bigger than most of the previous theoretical studies, except the recent one
of Ref.~\cite{Escribano} 
using Pad\'e approximants   to analyze the data of Refs.~\cite{Cello_1991,CLEO_1998,BaBar_2011}, see
Table~\ref{tab:results}.

At present the data available for $\eta'\to\pi\pi\gamma$ are not good enough to
constrain the slope parameter $\alpha$ of Eq.~(\ref{alphadef}) sufficiently to repeat
the analysis from above also for the $\eta'$. However, as suggested by the existing data---{\it cf.\/} 
the lower panel of Figure~\ref{fig:overOmnes}---as well as by the fact that  both decays
$\eta\to\pi\pi\gamma$ and $\eta'\to\pi\pi\gamma$ have the same leading left-hand cut,
we may now {\em assume} that the value of $\alpha$ given in Eq.~(\ref{alphaval}) also applies
to radiative $\eta'$ decays. Then, the only thing that changes compared to the analysis above
is the pre-factor $\kappa_{\eta}$ in Eq.~(\ref{eq:slope}) which is replaced by $\kappa_{\eta'} \equiv {\e} A^{\eta'}_{\pi\pi\gamma} f_\pi^2/A^{\eta'}_{\gamma\gamma}$  where the ratio of amplitude factors $A^{\eta'}_{\pi\pi\gamma}$ and
$A^{\eta'}_{\gamma\gamma}$ follows from ratio of the square roots of the corresponding branching ratios. 
In this way we get
\begin{equation}
 b_{\eta'}^{(I=1)} =  \left(1.19  \ \mbox{}^{+0.10}_{-0.04 } \right)\text{GeV}^{-2}  \, ,
  \label{res_bp_isovector}
\end{equation}
where the theoretical uncertainty is estimated in the same way as in the $\eta$ case.
If again the isoscalar contribution is added,
the full slope of the transition form factor is given by
\begin{equation}
 b_{\eta'} =  \left(1.53 \ \mbox{}^{+0.15}_{-0.08}  \right)\text{GeV}^{-2} 
 \label{res_bp_full}
\end{equation}
where the central values for $b_{\eta'}^{(I=1)}$ and $b_{\eta'}^{(I=0)}$ were added and the
increase in the uncertainty comes from the isoscalar part.
The result (\ref{res_bp_full}) 
is compatible with all experimental results, especially
with the Pad\'e-approximants fit~\cite{Escribano}  to
the \cite{Cello_1991,CLEO_1998,BaBar_2011} data and
with the predictions of  \mbox{1-loop} ChPT   as well as  VMD,
see Table~\ref{tab:results_prime}.

As a test of  internal consistency, we evaluated the analogous sum rule to 
Eq.~(\ref{check_full}) also for the $\eta'$.
In fact, if the integral occurring in the $\eta'$ analog of  Eq.\,(\ref{check_isovec}), namely in the isovector part of
the sum rule, 
is again truncated at $1\,\text{GeV}^2$, the right-hand side of the total
$A^{\eta'}_{\gamma\gamma}$
sum rule, 
\begin{eqnarray}\nonumber
  A^{\eta'}_{\gamma\gamma} &=& A^{\eta' (I=1)}_{\gamma\gamma} +A^{\eta' (I=0)}_{\gamma\gamma} \\
&=&  A^{\eta' (I=1)}_{\gamma\gamma} +(w_{\eta'\omega\gamma} + w_{\eta'\phi\gamma})A^{\eta'}_{\gamma\gamma} \ , \label{check_full_p}
\end{eqnarray}
which also contains the model-dependent isoscalar term, 
underestimates the
left-hand one by
$(-2 \pm 7)\%$.
If the upper integration range is varied as in the analogous expression for the $\eta$,
then these numbers change to
 $(-2\pm 9)\%$.

%%%%%%%%%%%%%%%%%%%%%%%%%%%%%%%%%%%%%%%%%%%%%%%%%%%%%%%%%%%%%%%%%%%%%%%%%%%%%%%%%%%%%%%%%%%%%%%%%%%%%%%%%%%%%%%%%%%%

\section{Summary and discussion}
\label{sec:discussion}

In summary, we have derived a model-independent integral representation for the isovector contribution to the $\eta\to \gamma\gamma^*$ transition form factor at 
low energies and especially the corresponding slope parameter $b_\eta$.
The necessary input was taken directly from experimental data, namely
from the pion vector form factor, the tabulated 
branching ratios for the $\eta\to \gamma\gamma$ and
$\eta\to\pi^+\pi^-\gamma$ decays, and from the measured spectral shape of  the latter process,
 parametrized by just  one coefficient,  the slope parameter $\alpha$ as described in Ref.~\cite{Stollenwerk}
 and Eq.~(\ref{alphadef}). 

This was possible with the help of the machinery of dispersion theory, by utilizing the fact that
the pion vector form factor and the  $\eta\to \pi^+\pi^-\gamma$   (and $\eta'\to \pi^+\pi^-\gamma$) decay
amplitudes have the same right-hand cut---at least in the region below $1\,\text{GeV}^2$ for the invariant pion mass square, the region dominating the {\em once}-subtracted dispersion relation. As a consistency check we demonstrated that a related {\em un}-subtracted dispersion integral is saturated at 1 GeV$^2$.

The isoscalar contribution of the slope parameter $b_\eta$, modelled by a simple vector-meson-dominance
approximation, turned out to be smaller than the uncertainty of the calculation for the isovector part
in the $\eta$ case. In the $\eta'$ scenario, the isoscalar part was larger, but still of subleading nature.
In addition, the isoscalar contributions when added to the isovector ones helped in saturating 
the {\em un}-subtracted $\eta\to\gamma\gamma$ and $\eta'\to\gamma\gamma$ sum rules below $1\,\text{GeV}^2$ to an uncertainty better than $12\%$ and $9\%$, respectively.

Our final results for the slopes of the  $\eta$ transition form factor are 
$b_\eta^{(I=1)} = \left(2.09  \mbox{}^{+0.21}_{-0.11} \right)\text{GeV}^{-2}$
for the isovector contribution
and
$b_\eta = \left(2.05 \ \mbox{}^{+0.22}_{-0.10} \right)\text{GeV}^{-2}$
in total. 
In fact, the slope at the origin of the transition form factor following the lower edge of the (orange) band in Figure~\ref{fig:formfactor} 
corresponds to
our prediction for the lower bound on the slope parameter, {\it i.e.\/} $b_\eta \geq 1.95\,\text{GeV}^{-2}$.
This value  is compatible with all recent
experimental results, but bigger than most previous theoretical studies known to us.

The available data for the $\eta'\to \pi\pi\gamma$ spectral shape are not good enough
to allow for a compatible fit of the corresponding $\alpha$ parameter.
 However, the slope parameter $\alpha$ solely determined
from the high-precision $\eta\to\pi\pi\gamma$ data of Ref.~\cite{KLOE_2013}  also provided a good fit to
the available $\eta'\to\pi\pi\gamma$ spectral data---without any readjustment.
Therefore, we conjectured that 
the value of $\alpha$ determined
in $\eta\to \pi\pi\gamma$ also applies to $\eta'\to \pi\pi\gamma$.

Under this assumption and  the inclusion of the model-dependent but sub\-leading iso\-scalar contributions,
which were derived in the same way as for the $\eta$, the following results  apply 
for the slope parameter $b_{\eta'}$: the isovector contribution reads
$b_{\eta'}^{(I=1)} = \left(1.19  \ \mbox{}^{+0.10}_{-0.04} \right)\text{GeV}^{-2}$
while the total result is
$b_{\eta'} = \left(1.53 \ \mbox{}^{+0.15}_{-0.08} \right)\text{GeV}^{-2}$.
Our result for $b_{\eta'}$ is 
compatible with all known experimental data, which, however, are rather old, and---this time---also with
chiral perturbation theory truncated at 1-loop order and with VMD. 

In case of the $\eta$ slope parameter there seems to be some tension between the  values determined from 
experimental data and the ones calculated by the dispersion integral. However, in this
context it should be stressed that the empirical slopes have been extracted from experimental data usually
with the help of monopole fits. Those have typically a larger curvature than our main result---{\it cf.\/} (orange) band in Figure~\ref{fig:formfactor}.
Thus, when the slope at $Q^2=0$ is extracted  from a monopole fit of the data, say well above 
the $\mu^+\mu^-$ threshold, the results are characteristically  smaller than those derived from the functional form of our final result. 

The formalism presented here allows us to disentangle the effects on the form factor slope emerging 
from the $\pi\pi$-interaction,
which are universal, from those
of the production vertex, which are reaction specific.  Our results show that the production vertex itself, whose effect is encoded in the parameter
$\alpha$, contributes to about 20\% of the slope of the transition form factor, while the bulk is provided by the $\pi\pi$ intermediate
state, which might be viewed as coming from the pole of $\rho$-meson. Therefore, parametrizing the transition form factor as a single 
monopole term, which suggests that the mass scale relevant for $\eta\to \gamma\gamma^*$ is entirely controlled by a single, reaction-dependent scale, is 
misleading, since the actual shape of the form factor emerges from the interplay of two scales.

%%%%%%%%%%%%%%%%%%%%%%%%%%%%%%%%%%%%%%%%%%%%%%%%%%%%%%%%%%%%%%%%%%%%%%%%%%%%%%%%%%%%%%%%%%%%%%%%%%%%%%%%%%%%%%%%%%%%

\begin{acknowledgements}
We would like to thank Martin Hoferichter,  Bastian Kubis, Franz Niecknig, Stefan Leupold and Simon
Eidelman
for helpful discussions and advice.
We are grateful to Camilla Di Donato for providing the KLOE data and to Marc Unverzagt for help
in connection with Figure~\ref{fig:formfactor}.
This work is supported in part by the DFG and the NSFC through
funds provided to the Sino-German CRC~110 ``Symmetries and
the Emergence of Structure in QCD'', and by the European
Com\-muni\-ty-Research  Infrastructure  Integrating  Activity  ``Study of
Strong\-ly Interacting Matter'' (acronym Hadron\-Physics3).
\end{acknowledgements}

%%%%%%%%%%%%%%%%%%%%%%%%%%%%%%%%%%%%%%%%%%%%%%%%%%%%%%%%%%%%%%%%%%%%%%%%%%%%%%%%%%%%%%%%%%%%%%%%%%%%%%%%%%%%%%%%%%%%

\appendix

\section{Comparison with the vector meson dominance approximation}  
\label{app: VMD}

It is instructive to compare Eq.~(\ref{eq:slope}) with what can be derived
from a simple realization of vector meson dominance (VMD).
For this purpose, we may write
\begin{eqnarray}
 P(s)^{\rm VMD}          &=&1 \qquad {(\text{{\it i.e.\/}:}\ \ \alpha =0)}\ , \label{eq:Pvmd} \\
 F_V(s)^{\rm VMD}      &=&  \frac{m_\rho^2}{m_\rho^2 - s - i m_\rho
   \Gamma_\rho(s)} \ , \label{eq:FVvmd} \\
 \kappa_\eta^{\rm VMD} &=&1 \qquad (\text{{\it i.e.\/}:}\ \
 A_{\gamma\gamma}^\eta =  {\e}A_{\pi\pi\gamma}^\eta {f_\pi^2}) \
 . \label{eq:Kvmd}
\end{eqnarray}
To proceed  we use $\pi \delta(x-x_0) = {\lim}_{\epsilon \to 0} \ \frac{\epsilon}{(x-x_0)^2 + \epsilon^2}$ 
to  approximate the form factor square 
with the help of the substitution  $\epsilon\equiv m_\rho \Gamma_\rho(s')$ as follows:
\begin{eqnarray}
 \lefteqn{
 \left | F_V(s^\prime)^{\rm VMD} \right|^2 
 =      \frac{m_\rho^4}{(m_\rho^2 - s^\prime)^2 + m_\rho^2 
 \Gamma_\rho(s')^2}
 }
 \nonumber \\ 
 &&=    \frac{m_\rho^3}{\Gamma_\rho(s')} \,\frac{\epsilon}{(m_\rho^2 - s^\prime)^2 + \epsilon^2}  
                                      \approx  
                                        \frac{m_\rho^3}{\Gamma_\rho(s')}\, \pi \delta\! \left(\! s^\prime - m_\rho^2 \right)\,.
  \label{eq:FVvmdsquared}
\end{eqnarray}

\noindent
Inserting (\ref{eq:Pvmd}),  (\ref{eq:Kvmd}) and  (\ref{eq:FVvmdsquared})  into Eq.~(\ref{eq:slope}) yields
\begin{eqnarray}
 b^{(I=1)}_{\eta\, \text{VMD}} &\approx& \frac{1}{96 \pi^2f_\pi^2} \ \int_{4 m_\pi^2}^\infty
 \frac{ds^\prime}{s^\prime}  \ \sigma_\pi(s^\prime)^3 \,\frac{m_\rho^3}{\Gamma_\rho(s')} \,\pi \delta\! \left(\!s^\prime - m_\rho^2 \right) \nonumber \\
                  &=&    \frac{1}{96 \pi f_\pi^2} \ \frac{m_\rho}{\Gamma_\rho
   (m_\rho^2)} \left( \sigma_\pi (m_\rho^2)\right)^3 \,.
    \label{eq:b_approx}
\end{eqnarray}
We may now employ the explicit form of the width of $\rho$,
\begin{equation}
 \Gamma_\rho (m_\rho^2)= \frac{1}{48 \pi} \, g_{\rho\pi\pi}^2 \, m_\rho  \left(\sigma_\pi(m_\rho^2)\right)^3\ ,
\end{equation}
namely the (spin-averaged) standard two-body decay formula~\cite{pdg:2012}
with $\mathcal M (\rho^0 \to \pi^+(p_1) \pi^- (p_2))
= g_{\rho\pi\pi} |\vec p_1-\vec p_2|$ as amplitude and  $g_{\rho\pi\pi}$ as coupling constant.
In this way we get
\begin{equation}
 b^{(I=1)}_{\eta\, \text{VMD}} \approx \frac{1}{2f_\pi^2 g_{\rho\pi\pi}^2} \approx \frac{1}{m_\rho^2} \ ,
\end{equation}
where in the last step the KSFR relation $g_{\rho\pi\pi}^2 \approx m_\rho^2/(2 f_\pi^2)$ was 
applied~\cite{KSFRa,KSFRb}.

Thus our formalism naturally matches onto the VMD approximation---see, {\it e.g.\/}, 
Refs.~\cite{Landsberg,UGM_VMD} for reviews and \cite{Benayoun_EPJC55,Benayoun_EPJC65,Benayoun_EPJC17b,Benayoun_EPJC72,Carla_2010,Carla_2012} for recent updates---once the corresponding
expressions for the various ingredients are imposed. If we had kept the empirical value 
$\kappa_\eta=0.566\pm 0.006$ and
inserted the linear polynomial $P(s') = 1 + \alpha s'$ instead of 
Eq.~(\ref{eq:Pvmd}) into the integral of Eq.~(\ref{eq:b_approx}), we would have got the modified approximation 
 \begin{equation}
b^{(I=1)}_{\eta\,\text{mod.VMD}} \approx \frac{\kappa_\eta}{m_\rho^2} (1 + \alpha m_\rho^2) 
\end{equation}
for the isovector part of the slope.
In this case the VMD result would be enlarged by a  factor $1+\alpha m_\rho^2 \approx 1.79\pm 0.08$, namely
by the linear polynomial $P(s')$ evaluated at $s'= m_\rho^2$ with $\alpha$ as in Eq.~(\ref{alphaval}), while
the empirical prefactor $\kappa_\eta$ would nearly counterbalance this result, such that approximately the original VMD result, 
\begin{equation}
b^{(I=1)}_{\eta\,\text{mod.VMD}} \approx (1.02\pm0.05)/m_\rho^{2}\approx (1.69\pm 0.08)\,\text{GeV}^{-2}\,, 
\end{equation}
reemerges. The latter is---as expected---markedly smaller 
than our prediction (\ref{res_b_isovector})  from  the dispersion integral (\ref{eq:slope}).

%%%%%%%%%%%%%%%%%%%%%%%%%%%%%%%%%%%%%%%%%%%%%%%%%%%%%%%%%%%%%%%%%%%%%%%%%%%%%%%%%%%%%%%%%%%%%%%%%%%%%%%%%%%%%%%%%%%%

% BibTeX users please use one of
%\bibliographystyle{spbasic}      % basic style, author-year citations
%\bibliographystyle{spmpsci}      % mathematics and physical sciences
%\bibliographystyle{spphys}       % APS-like style for physics
%\bibliography{}   % name your BibTeX data base

\begin{thebibliography}{99}

\bibitem{krakau}
E.~Czerwinski, S.~Eidelman, C.~Hanhart, B.~Kubis, A.~Kup\'s\'c, S.~Leupold, P.\,Moskal, S.\,Schadmand,
  %``MesonNet Workshop on Meson Transition Form Factors,''
  arXiv:1207.6556 [hep-ph].
  %%CITATION = ARXIV:1207.6556;%%

\bibitem{trento}
  H.~Czy\.z {et al.},
  %``Constraining the Hadronic Contributions to the Muon Anomalous Magnetic Moment,''
  arXiv:1306.2045 [hep-ph].
  %%CITATION = ARXIV:1306.2045;%%

\bibitem{Stollenwerk} 
  F.~Stollenwerk, C.~Hanhart, A.~Kup{\'s}{\'c}, \mbox{U.-G.}~Mei{\ss}ner, A.~Wirzba,
  %``Model-independent approach to eta -> pi+ pi- gamma and eta' -> pi+ pi- gamma,''
  Phys. Lett. B  \textbf{707}, 184 (2012)
  [arXiv:1108.2419 [nucl-th]]
  
\bibitem{pdg:2012}
  J.~Beringer et al. (Particle Data Group Collaboration),
  %``Review of Particle Physics (RPP),''
  Phys. Rev. D \textbf{86},  010001 (2012)
  and 2013 partial update for the 2014 edition

\bibitem{WASAeta2pipiga}
 P.~Adlarson {et al.}  (WASA-at-COSY Collaboration),
  %``Exclusive Measurement of the $\eta \to \pi^+ \pi^- \gamma$ Decay,''
  Phys. Lett. B \textbf{707}, 243 (2012)
  [arXiv:1107.5277 [nucl-ex]]
  %%CITATION = ARXIV:1107.5277;%% 

\bibitem{Crystal_Barrel}
 A.~Abele  {et al.}  (Crystal Barrel Collaboration),
  %``Measurement of the decay distribution of eta-prime --> pi+ pi- pi- gamma and evidence for the box anomaly,''
  Phys. Lett. B \textbf{402}, 195 (1997) 
  %%CITATION = PHLTA,B402,195;%%

\bibitem{KLOE_2013}
  D.~Babusci {et al.}  (KLOE/KLOE-2 Collaboration),
  %``Measurement of $\Gamma(\eta \to \pi^+\pi^-\gamma)/\Gamma(\eta \to \pi^+\pi^-\pi^0)$ with the KLOE Detector,''
  Phys. Lett. B \textbf{718}, 910  (2013) 
  [arXiv:1209.4611 [hep-ex]]
  %%CITATION = ARXIV:1209.4611;%%

\bibitem{madrid}
  R.~Garcia-Martin {et al.},
  %``The Pion-pion scattering amplitude. IV: Improved analysis with once subtracted Roy-like equations up to 1100 MeV,''
  Phys. Rev. D \textbf{83},  074004 (2011) 
  [arXiv:1102.2183 [hep-ph]]

\bibitem{newff}
  C.~Hanhart,
  %``A New Parameterization for the Pion Vector Form Factor,''
  Phys. Lett. B \textbf{715}, 170 (2012) 
  [arXiv:1203.6839 [hep-ph]]

 \bibitem{belle}
  M.~Fujikawa  {et al.}  (Belle Collaboration),
  %``High-Statistics Study of the tau- ---> pi- pi0 nu(tau) Decay,''
  Phys. Rev. D \textbf{78}, 072006 (2008)
  [arXiv:0805.3773 [hep-ex]]
  %%CITATION = ARXIV:0805.3773;%%

 \bibitem{BKM_91}
   V.~Bernard, N.~Kaiser, U.-G.~Mei{\ss}ner,
  %``pi eta scattering in QCD,''
  Phys. Rev. D \textbf{44},  3698 (1991) 
 
\bibitem{bastian}
   B.~Kubis, S.P.~Schneider,
   %``The Cusp effect in eta-prime ---> eta pi pi decays,''
   Eur. Phys. J.  C \textbf{62},  511 (2009) 
 
\bibitem{BESIII}
 D.M.~Asner  {et al.}, 
 %``Physics at BES-III,''
 Int.  J. Mod. Phys. A \textbf{24}, S1 (2009)
 [arXiv:0809.1869 [hep-ex]]
 
\bibitem{Primenet}
P.~Adlarson, M.~Amaryan, M.~Bashkanov, F.~Bergmann, M.~Berlowski, J.~Bijnens, L.C.~Balkestahl, D.~Coderre  
{et al.},
  %``Proceedings of the second International PrimeNet Workshop,''
  arXiv:1204.5509 [nucl-ex]
  %%CITATION = ARXIV:1204.5509;%%

 \bibitem{freds}
  F.~Jegerlehner, R.~Szafron,
  %``$\rho^0 - \gamma$ mixing in the neutral channel pion form factor $F_{\pi}^{e}$ and its role in comparing 
  % $e^+  e^-$ with $\tau$ spectral functions,''
  Eur. Phys. J.  C \textbf{71},  1632 (2011)

 \bibitem{Landsberg} 
  L.G.~Landsberg,
  %``Electromagnetic Decays of Light Mesons,''
  Phys. Rep.  \textbf{128}, 301 (1985) 
  %%CITATION = PRPLC,128,301;%%

 \bibitem{Bijnens_1990}
  J.~Bijnens, A.~Bramon, F.~Cornet,
  %``Three Pseudoscalar Photon Interactions In Chiral Perturbation Theory,''
  Phys. Lett.  B  \textbf{237}, 488 (1990) 
  %%CITATION = PHLTA,B237,488;%%

\bibitem{barry}
B.R.~Holstein, 
Phys.  Scripta T \textbf{99}, 55 (2002)
%{{\tt arXiv:hep-ph/0112150}}.
%%CITATION = HEP-PH/0112150;%%.

\bibitem{Benayoun_EPJC55}
  M.~Benayoun, P.~David, L.~DelBuono, O.~Leitner, H.B.~O'Connell,
  %``The Dipion Mass Spectrum In e+ e- Annihilation and tau Decay: A Dynamical (rho, omega, phi) Mixing Approach,''
  Eur. Phys.  J.  C \textbf{55}, 199  (2008) 
  [arXiv:0711.4482 [hep-ph]]

\bibitem{Benayoun_EPJC65}
  M.~Benayoun, P.~David, L.~DelBuono, O.~Leitner,
  %``A Global Treatment Of VMD Physics Up To The phi: I. e+ e- Annihilations, 
  % Anomalies And Vector Meson Partial Widths,''
  Eur.  Phys. J.  C \textbf{65}, 211 (2010)
  [arXiv:0907.4047 [hep-ph]]
  %%CITATION = ARXIV:0907.4047;%%

\bibitem{Benayoun_EPJC17b}
  M.~Benayoun, L.~DelBuono, H.B.~O'Connell,
  %``VMD, the WZW Lagrangian and ChPT: The Third mixing angle,''
  Eur. Phys. J. C \textbf{17}, 593 (2000)
  [hep-ph/9905350]
  %%CITATION = HEP-PH/9905350;%%

 \bibitem{Benayoun_EPJC72}
  M.~Benayoun, P.~David, L.~DelBuono, F.~Jegerlehner,
  %``Upgraded Breaking Of The HLS Model: A Full Solution to the $\tau^-e^+e^-$ 
  % and $\phi$ Decay Issues And Its Consequences On g-2 VMD Estimates,''
  Eur. Phys. J. C \textbf{72}, 1848 (2012)
  [arXiv:1106.1315 [hep-ph]]
  %%CITATION = ARXIV:1106.1315;%%

\bibitem{VEPP-2M} 
  M.N.~Achasov, S.E.~Baru, A.V.~Bozhenok, A.D.~Bukin, D.A.~Bukin, S.V.~Burdin, T.V.~Dimova, 
  S.I.~Dolinsky {et al.},
  %``Study of QED processes e+ e- ---> e+ e- gamma, e+ e- gamma gamma with the SND detector at VEPP-2M,''
  Eur. Phys. J. C \textbf{12}, 369 (2000)
  [hep-ex/9908068]
  %%CITATION = HEP-EX/9908068;%%

\bibitem{CMD2-epem} 
  R.~R.~Akhmetshin {et al.}  (CMD2 Collaboration),
  %``Study of the processes e+ e- ---> eta gamma, pi0 gamma ---> 3 gamma in the c.m. energy range 600-MeV to 1380-MeV at CMD-2,''
  Phys.  Lett. B \textbf{605}, 26 (2005)
  [hep-ex/0409030]
  %%CITATION = HEP-EX/0409030;%%

\bibitem{SND-epem} 
  M.N.~Achasov, K.I.~Beloborodov, A.V.~Berdyugin, A.G.~Bogdanchikov, A.D.~Bukin, D.A.~Bukin, T.V.~Dimova {et al.},
  %``Reanalysis of the e+ e- --->eta gamma reaction cross section,''
  Phys. Rev. D \textbf{76}, 077101 (2007)
  [arXiv:0709.1007 [hep-ex]]
  %%CITATION = ARXIV:0709.1007;%%

 \bibitem{BL}
  S.J. Brodsky, G.P. Lepage,
  %``Large Angle Two Photon Exclusive Channels in Quantum Chromodynamics,''
  Phys.  Rev.  D \textbf{24}, 1808 (1981)
  %%CITATION = PHRVA,D24,1808;%%

 \bibitem{Brodsky_2011}
 S.J. Brodsky, F.-G.~Cao, G.F. de Teramond,
  %``Evolved QCD predictions for the meson-photon transition form factors,''
  Phys. Rev. D \textbf{84}, 033001  (2011) 
  [arXiv:1104.3364 [hep-ph]]
  %%CITATION = ARXIV:1104.3364;%%

 \bibitem{Bijnens_Pers}
 J.~Bijnens, F.~Persson,
  %``Effects of different form-factors in meson photon photon transitions and the muon anomalous magnetic moment,''
  hep-ph/0106130
  %%CITATION = HEP-PH/0106130;%%

 \bibitem{Schneider} 
 S.P.~Schneider, B.~Kubis, F.~Niecknig,
 %``The $\omega -> \pi^0 \gamma^*$ and $\phi -> \pi^0 \gamma^*$ transition form factors in dispersion theory,''
 Phys. Rev. D \textbf{86}, 054013 (2012)
 [arXiv:1206.3098 [hep-ph]]
 %%CITATION = ARXIV:1206.3098;%%

 \bibitem{Bastian_Martin}
  M.~Hoferichter, B.~Kubis, D.~Sakkas,
  %``Extracting the chiral anomaly from gamma pi --> pi pi,''
  Phys. Rev. D \textbf{86}, 116009 (2012)
  [arXiv:1210.6793 [hep-ph]]
  %%CITATION = ARXIV:1210.6793;%%

 \bibitem{NA60_2011}
 G.~Usai (NA60 Collaboration),
 %``Low mass dimuon production in proton-nucleus collisions at 400-GeV/c,''
 Nucl. Phys. A \textbf{855}, 189 (2011)
 %%CITATION = NUPHA,A855,189;%%

 \bibitem{CB_TAPS_2011}
  H.~Berghauser, V.~Metag, A.~Starostin, P.~Aguar-Bartolome, L.K.~Akasoy,
   J.R.M.~Annand, H.J.~Arends, K.~Bantawa {et al.},
  %``Determination of the eta-transition form factor in the 
  % gamma p ---> p eta ---> p   gamma e+ e- reaction,''
  Phys. Lett. B {\bf 701}, 562 (2011)
  %%CITATION = PHLTA,B701,562;%%

 \bibitem{Escribano} R.~Escribano, P.~Masjuan, P.~Sanchez-Puertas,
  %``\eta and \eta' transition form factors from rational approximants,''
  arXiv:1307.2061v2 [hep-ph]

\bibitem{Cello_1991}
  H.J.~Behrend {et al.} (CELLO Collaboration),
  %``A Measurement of the pi0, eta and eta-prime electromagnetic form-factors,''
  Z. Phys. C \textbf{49}, 401 (1991) 
  %%CITATION = ZEPYA,C49,401;%%

\bibitem{CLEO_1998} 
J.~Gronberg {et al.}  (CLEO Collaboration),
  %``Measurements of the meson - photon transition form-factors of light pseudoscalar mesons at large momentum transfer,''
  Phys. Rev. D \textbf{57}, 33 (1998)
  [hep-ex/9707031]
  %%CITATION = HEP-EX/9707031;%%
  
 \bibitem{BaBar_2011} 
P.~del Amo Sanchez {et al.}  (BaBar Collaboration),
  %``Measurement of the $\gamma \gamma^* --> \eta$ and $\gamma \gamma* --> \eta'$ transition form factors,''
  Phys. Rev. D  \textbf{84}, 052001 (2011)
  [arXiv:1101.1142 [hep-ex]]
  %%CITATION = ARXIV:1101.1142;%%

 \bibitem{Ametller_1992}
  L.~Ametller, J.~Bijnens, A.~Bramon, F.~Cornet,
  %``Transition form-factors in pi0, eta and eta and eta-prime couplings to gamma gamma,''
  Phys. Rev. D \textbf{45}, 986 (1992)
  %%CITATION = PHRVA,D45,986;%

 \bibitem{SND_2001}
 M.N.~Achasov, V.M.~Aulchenko, K.I.~Beloborodov, A.V.~Berdyugin, A.G.~Bogdanchikov,
  A.V.~Bozhenok, A.D.~Bukin, D.A.~Bukin {et al.},
   %``Study of Conversion Decays phi --> eta e+ e- and eta --> gamma e+ e- in the 
   %  Experiment with SND Detector at the VEPP-2M Collider,''
   Phys.  Lett.  B \textbf{ 504} 275, (2001) 

 \bibitem{Lepton_G_1980}
  R.I.~Dzhelyadin, S.V.~Golovkin, V.A.~Kachanov, A.S.~Konstantinov,
   V.F.~Konstantinov, V.P.~Kubarovsky, A.V.~Kulik, L.G.~Landsberg {et al.},
  %``INVESTIGATION OF eta MESON ELECTROMAGNETIC STRUCTURE IN eta -> mu+ mu- gamma DECAY,''
  Phys. Lett. B \textbf{94}, 548 (1980) 
   [Sov. J. Nucl. Phys.  \textbf{32}, 516 (1980)]
   [Yad. Fiz.  \textbf{32}, 998 (1980) 998]
  %%CITATION = PHLTA,B94,548;%%

 \bibitem{TPC2gamma}
  H.~Aihara {et al.} (TPC/Two Gamma Collaboration),
  %``Investigation of the electromagnetic structure of $\eta$ and $\eta^\prime$ mesons by two photon interactions,''
  Phys. Rev. Lett.  \textbf{64}, 172 (1990) 
  %%CITATION = PRLTA,64,172;%%  

 \bibitem{NA60_2009}
  R.~Arnaldi  {et al.}  (NA60 Collaboration),
  %``Study of the electromagnetic transition form-factors in 
  % eta --> mu+ mu- gamma and omega ---> mu+ mu- pi0 decays with NA60,''
  Phys. Lett. B \textbf{677}, 260 (2009)
  [arXiv:0902.2547 [hep-ph]]
  %%CITATION = ARXIV:0902.2547;%%

 \bibitem{QL1}
  A.~Bramon, E.~Masso,
  %``Q**2 Duality For Electromagnetic Form-factors Of Mesons,''
  Phys. Lett.  B \textbf{104}, 311 (1981)
  %%CITATION = PHLTA,B104,311;%%

\bibitem{QL2}
  L.~Ametller, L.~Bergstrom, A.~Bramon, E.~Masso,
  %``The Quark Triangle: Application To Pion And Eta Decays,''
  Nucl. Phys. B \textbf{228}, 301 (1983)
 %%CITATION = NUPHA,B228,301;%%

\bibitem{QL3}
  A.~Pich, J.~Bernabeu,
  %``Rare Decay Modes Of The Neutral Pion,''
  Z. Phys. C \textbf{22}, 197 (1984) 
  %%CITATION = ZEPYA,C22,197;%%

\bibitem{Lepton_G_1979}
R.I.~Dzhelyadin, S.V.~Golovkin, M.V.~Gritsuk, V.A.~Kachanov, D.B.~Kakauridze, A.S.~Konstantinov, V.F.~Konstantinov, V.P.~Kubarovsky {et al.},
  %``OBSERVATION OF eta-prime ---> mu+ mu- gamma DECAY,''
  Phys. Lett. B \textbf{88}, 379 (1979) 
   [JETP Lett. \textbf{30}, 359 (1979)]
  %%CITATION = PHLTA,B88,379;%%
  
\bibitem{KSFRa}
  K.~Kawarabayashi, M.~Suzuki,
  %``Partially conserved axial vector current and the decays of vector mesons,''
  Phys. Rev. Lett.  \textbf{16}, 255 (1966)
  %%CITATION = PRLTA,16,255;%%

\bibitem{KSFRb}
Riazuddin, Fayyazuddin,
  %``Algebra of current components and decay widths of rho and K* mesons,''
  Phys. Rev.  \textbf{147}, 1071  (1966)
  %%CITATION = PHRVA,147,1071;%%

\bibitem{UGM_VMD} 
  U.-G.~Mei{\ss}ner,
  %``Low-Energy Hadron Physics from Effective Chiral Lagrangians with Vector Mesons,''
  Phys. Rep.  \textbf{161}, 213 (1988)
  %%CITATION = PRPLC,161,213;%%

\bibitem{Carla_2010}
C.~Terschl\"usen, S.~Leupold,
 %``Electromagnetic transition form factors of light vector mesons,''
 Phys. Lett. B  \textbf{691}, 191 (2010)
 [arXiv:1003.1030 [hep-ph]]
 %%CITATION = ARXIV:1003.1030;%%

\bibitem{Carla_2012}
 C.~Terschl\"usen, S.~Leupold, M.F.M.~Lutz,
 %``Electromagnetic Transitions in an Effective Chiral Lagrangian with the eta-prime and Light Vector Mesons,''
 arXiv:1204.4125v2 [hep-ph]
 %%CITATION = ARXIV:1204.4125;%%

\end{thebibliography}

% Non-BibTeX users please use

%%%%%%%%%%%%%%%%%%%%%%%%%%%%%%%%%%%%%%%%%%%%%%%%%%%%%%%%%%%%%%%%%%%%%%%%%%%%%%%%%%%%%%%%%%%%%%%%%%%%%%%%%%%%%%%%%%%%

\end{document}